\documentclass[aps,prl,twocolumn, superscriptaddress]{revtex4-2}

\usepackage{graphicx}
\usepackage{amssymb}
\usepackage{amsmath}
\usepackage{color}
\usepackage{graphicx}
\usepackage{dcolumn}
\usepackage{epstopdf}
\usepackage{bm}
\usepackage{natbib}
\usepackage{tikz}
\usepackage{siunitx}
\usepackage[colorlinks=true,citecolor=blue]{hyperref}
\usepackage{soul}
\usepackage{braket}

\newcommand{\orcid}[1]{\href{https://orcid.org/#1}{\includegraphics[width=8pt]{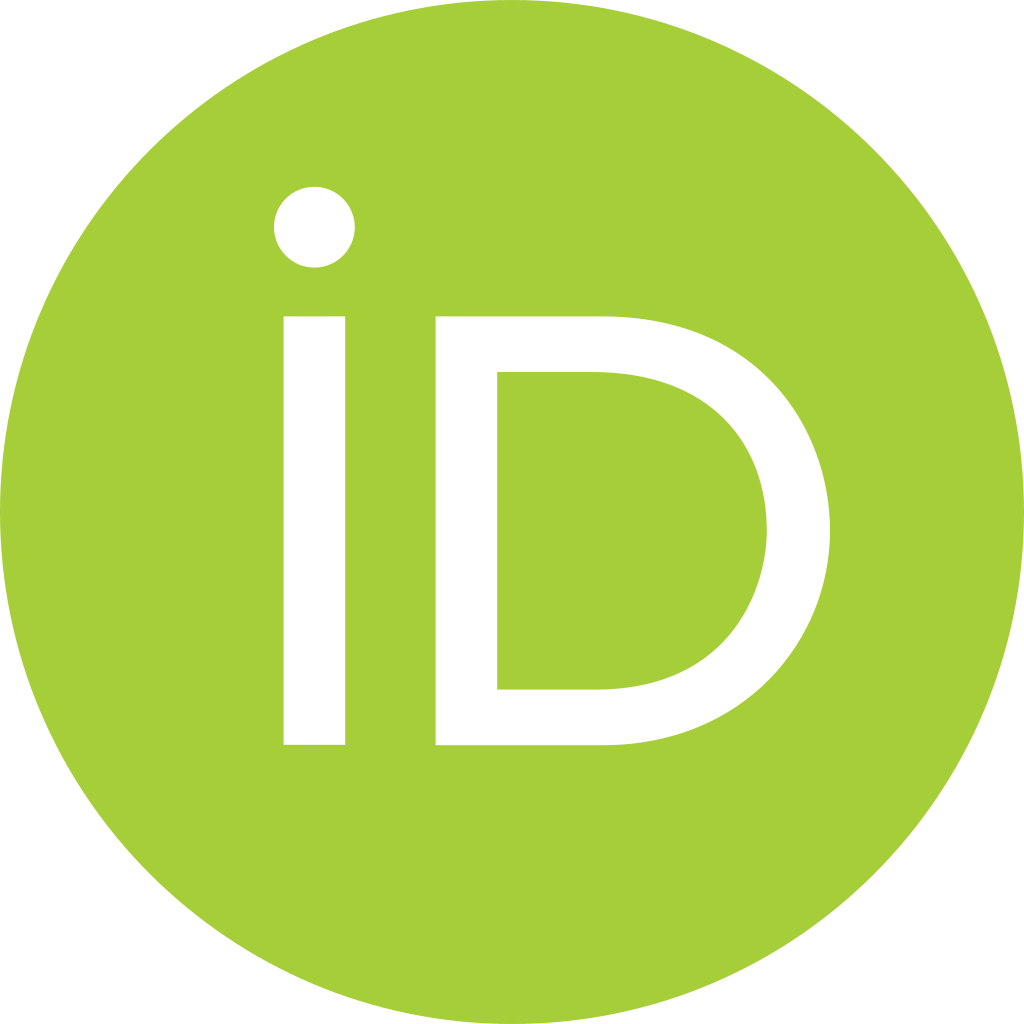}}}
\newcommand{\edit}[1]{{\color{black} #1}}

\begin{document}

\title{Photonic Mode Description of the Jaynes-Cummings Hamiltonian States}

\author{V. F. Maisi \orcid{0000-0003-4723-7091}}
\email{ville.maisi@ftf.lth.se}
\affiliation{NanoLund and Solid State Physics, Lund University, Box 118, 22100 Lund, Sweden}

\begin{abstract}
    Jaynes-Cummings Hamiltonian provides the elemental description of a two-level system interacting with a photonic mode. In this work, we review the Jaynes-Cummings formalism for an electronic two-level system. \edit{The purpose is to summarize and connect together the key theory concepts relevant for the transmission response probed typically in the experiments. The review provides a photonic mode description of the response with analytical results for several limiting cases. These expressions are useful for gaining understanding and intuition on what influences the measured response. We also} connect \edit{the photonic response} to several key concepts in the field: the dispersive shift \edit{in qubit readout, the} quantum capacitance \edit{of semiconductor nanostructures} and the effective dissipation rate \edit{of the two-level system}. 
\end{abstract}

\date{\today}
\maketitle

\section{Introduction}

Jaynes-Cummings (JC) Hamiltonian has a pivotal role in quantum physics. It is widely used to describe the light-matter interaction of atoms and optical cavities~\cite{Kimble1998, Haroche2006, Thompson1992, Brune1996, Reithmaier2004, Yoshie2004, Kavokin2015, Kavokin2017, Zeb2018, Delteil2019}, as well as many solid-state quantum devices~\cite{Clerk2020, Blais2021, Wallraff2004, Mi2017, Stockklauser2017, Burkard2020, Ranni2024}. The JC Hamiltonian describes how a transition from one quantized state to another interacts coherently with a harmonic photonic mode, leading to the well-known Rabi splitting and Rabi oscillations in the strong-coupling regime when the coupling strength exceeds the losses in the system. The hybridization of the electronic and photonic states is also a generally known outcome of the coherent coupling\edit{. T}he JC Hamiltonian together with Lindblad equation\edit{~\cite{Franke1976, Gorini1976, Lindblad1976}} and input-output formalism\edit{~\cite{Gardiner1984}} predicts that the coherence of the hybridized states is \edit{given by} a weighted sum of the electronic and photonic contributions.

One of the most common ways to \edit{study} systems following the JC Hamiltonian is to probe the transmission and reflection coefficient via couplings to the photonic mode. \edit{The hybridized modes result in resonant peaks to the response, making this approach efficient to gain understanding of the JC system. It is typically of specific interest to study the properties of the electronic part and its influence to the overall response. Among the first few key tasks in the experiment is typically t}o determine the coupling strengths and \edit{the} coherence of the electronic \edit{system. These measurements typically focus on the spectral properties such as the Rabi splitting and the broadening of the resonance peaks}~\cite{Thompson1992, Brune1996, Wallraff2004, Reithmaier2004, Diniz2011, Stockklauser2017, Mi2017, Zeb2018, Delteil2019}. \edit{The peak transmission is also a key property in the experiments as it depends on the couplings and decoherence rates of the system. It is also of practical importance as it describes how strong the measured signal is. If the peak transmission is too low, the resonance peak cannot be resolved.} 

\edit{In this work, we} review the JC formalism for a single two-level system (TLS) \edit{at low probe power and focus on the above important aspects. We consider a} double quantum dot structure as an exemplary structure for the TLS~\cite{Childress2004}. Most of the \edit{considered concepts}, however, are applicable to other systems as well,  for example, for superconducting charge qubits and transmons~\cite{Blais2004, Wallraff2004,Kjaergaard2020}. Here, mainly the coupling strengths and the precise interaction mechanism vary from system to system. 

\edit{We first summarize the derivation of the peak transmission, linewidth} and the couplings of the \edit{hybridized} modes at low probe power. The derivation yields analytic expressions \edit{connecting all of these quantities together and to the underlying} bare coupling rates and losses. \edit{Each} hybridized state can be perceived \edit{here as its own photonic mode} analogous to \edit{the} bare photonic resonator mode. \edit{This photonic mode description provides} an intuitive picture of the modes consisting of a "visible" photonic part that interacts with the input/output lines and a "hidden" electronic part that does not connect directly to the probing signals. 

\edit{We then illustrate the use of these results with two exemplary cases. In the first case, we consider a usual experimental situation where the electronic part has a high decoherence rate such that the peak transmission decreases with decreasing hybridization. We contrast this case with another one where the two-level system has lower decoherence rate than the photonic mode. For this case,} the \edit{peak transmission} remains strongly visible in the response despite \edit{of its linewidth and the mode} coupling decreas\edit{ing} significantly. The analytic expressions \edit{summarized} in this review describe \edit{these effects} in terms of changes in the effective input/output coupling and the internal losses of the mode. \edit{These exemplify how the photonic mode description provides} a simple and useful tool for understanding the features seen in the experimental response \edit{and interlinks them to the underlying properties of the TLS}. 

\edit{The review continues by connecting the transmission response and the photonic mode description further to several key concepts in the field. We discuss} how the \edit{transmission response} contains the dispersive \edit{frequency} shift induced to the resonator by the TLS~\cite{Blais2004, Boissonneault2009}. This \edit{constitutes the} usual way of performing the readout of the \edit{two-level system, i.e. the qubit}, as the shift depends on whether the TLS is in the ground or excited state~\cite{Kjaergaard2020}. \edit{ Furthermore, in the weak coupling limit, we make the link to so-called quantum capacitance which is a key concept for semiconducting nanostructures~\cite{Luryi1988, Petersson2010, Gonzalez-Zalba2016, Chatterjee2018, Vigneau2023}. In the weak coupling limit,} the response consists of only one mode close to the bare resonator frequency, and the TLS gives rise to \edit{a weak shift in the resonance frequency and an increase of the linewidth. The photonic mode description allows for showing how the resonance frequency shift arises from the quantum capacitance and the linewidth increase on the other hand from the underlying dissipation that the TLS induces to the mode. }

\begin{figure}[t]
    \centering
    \includegraphics[width=0.48\textwidth]{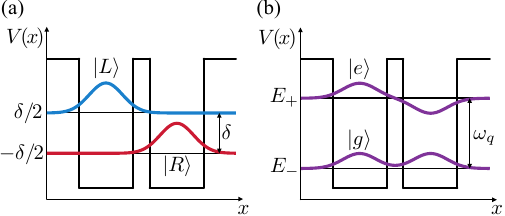}
    \caption{(a) The DQD states illustrated with the help of a 1-dimensional potential well $V(x)$. The left well has a discrete energy state $\ket{L}$ shown in blue and the right well the state $\ket{R}$ shown in red. The energy difference of these bare position states is $\delta$. (b) The hybridized states $\ket{g}$ and $\ket{e}$ resulting in the inter-dot tunnel coupling $t$ coupling the states $\ket{L}$ and $\ket{R}$ for the fully hybridized case $\delta = 0$.}
    \label{fig:DQDpot}
\end{figure}

\section{Double quantum dot as a charge qubit}
\edit{The TLS in the JC Hamiltonian consists of two quantized energy levels. The lower one of them is the ground state $\ket{g}$ of the TLS with energy $E_g$ and the higher one the excited state $\ket{e}$ with energy $E_e$. In the JC Hamiltonian, these states are considered to be coherent and interact coherently with the harmonic photonic mode. In the context of solid-state devices, predominantly considered in this review, such a coherent TLS is called a qubit as it constitutes the elemental component holding quantum information. The terms TLS and qubit in this review are hence interchangeable.

The above properties are general for any TLS and the material presented in this review is made based on these general concepts. It is, however illustrative to consider a specific example system. As the example system, we consider so-called double quantum dot (DQD) system. It is o}ne of the most common ways to form a charge qubit in a semiconducting nanostructure~\cite{Childress2004, Frey2012, Mi2017, Stockklauser2017}. \edit{This example system allows us to see how the underlying electronic states hybridize and form the ground and excited state, as well as how the underlying details of the specific system influence on the light-matter coupling.

Quantum dots (QDs) are nano-sized semiconductor structures. A quantum dot forces electrons to localize in such a small space that they show quantum effects. An electron travels bound to a QD travels back and forth in it reflecting from the boundaries. This movement back and forth results in the wavefunction of the electron to form a standing wave such that only certain wavelengths are allowed. As the wavelength is connected to the energy of the electron, this leads to the quantization of the allowed energies inside the QD and the formation of the discrete quantized energy levels.

The qubit in the DQD forms so that each dot has only one energy level relevant for the system. The quantization energy is so high that the other higher energy states never get occupied and the lower energy ones are always occupied with electrons stuck still to them at all times. This regime is achieved with small enough systems at sufficiently low temperatures so that the thermal energy $kT$ is much smaller than the energy level spacing in the QDs. In this case, the system can be illustrated with the diagram presented in Fig.~\ref{fig:DQDpot}. The two quantum dots are presented each with a one-dimensional potential well in the potential energy $V(x)$. The relevant quantized energy level $\ket{L}$ on the left dot is illustrated with the blue wavefunction that takes the form of a  standing wave with half wavelength in the QD. The red wavefunction shows a similar quantized energy level $\ket{R}$ in the right dot. Here we have chosen to illustrate the energy levels with}
the lowest bound states of a \edit{one}-dimensional well for simplicity. In practice, the states can be any bound states that are close-by in energy and have a single-electron that may reside in either of the states.
\edit{At low temperatures, adding or removing electrons from the DQD has a significant energy cost. This charging energy cost allows for having only a single electron residing in the $\ket{L}$ and $\ket{R}$ states. The energy cost depends on voltages applied to nearby gate electrodes. This allows to change the detuning $\delta$ between the energy levels which is directly proportional to the gate voltages. The changing energy cost and tuning with the gate voltages is thoroughly described in Ref.~\citealp{vanderWiel2002}. For the considerations below, w}e set the energy reference such that the left state $\ket{L}$ is at an energy $\delta/2$, and the right one $\ket{R}$ at $-\delta/2$ so that their energy difference is equal to $\delta$.

\edit{To have the qubit interacting efficiently with the photons, the two quantum dots need to be coupled to each other. Otherwise the electron is stuck to one of the dots and nothing happens in the system. The coupling of the two dots is achieved by making the energy barrier in $V(x)$ between the two dots to be thin enough. Then the electron may tunnel between the dots. This is described with the tunnel coupling} strength $t$\edit{. This results in the Hamiltonian of the DQD as}
\begin{equation}
\label{eq:DQDham}
\hat{H}_\mathrm{DQD} = \frac{\delta}{2}\left( \ket{L}\bra{L} - \ket{R}\bra{R} \right) + 
                        t \left( \ket{L}\bra{R} + \ket{R}\bra{L} \right).
\end{equation}
\edit{Here we have assumed that any other energy levels of the quantum dots are far away in energy so that they do not contribute. The $\ket{L}$ and $\ket{R}$ states are also treated as non-degenerate states. This assumption is typically made when describing charge qubit experiments~\cite{Childress2004, Frey2012, Mi2017, Stockklauser2017}. In other words, the spin degree of freedom is neglected in this case. The charging energies of the two dots are also assumed to be so high that the charge states are limited to the $\ket{L}$ and $\ket{R}$, i.e. one excess electron occupying either the left of the right dot.} Diagonalizing this Hamiltonian yields
\begin{equation}
\label{eq:DQDhamDiag}
\hat{H}_\mathrm{DQD} = E_\edit{e} \ket{e}\bra{e} + E_\edit{g} \ket{g}\bra{g},
\end{equation}
with $E_\edit{g} = -\sqrt{(\delta/2)^2 +t^2}$,  $E_\edit{e} = +\sqrt{(\delta/2)^2 +t^2}$, and the ground state $\ket{g} = c_-\ket{L} + c_+ \ket{R}$, and \edit{the} excited state $\ket{e} = c_+\ket{L} - c_- \ket{R}$. 
Here $c_\pm = \sqrt{(1 \pm \delta/\Omega)/2}$ are the so-called Hopfield coefficients~\cite{Hopfield1958, Kavokin2015, Kavokin2017, Ferreira2022}. They can be written equivalently as $c_+ = \cos(\phi/2)$ and $c_- = \sin(\phi/2)$ with the so-called mixing angle $\phi = \arctan(2t/\delta)$, with $0 < \phi < \pi$. The energy (difference) of the qubit is then $\hbar \omega_q = \edit{E_e - E_g} = \sqrt{\delta^2 + (2t)^2}$. \edit{Figure~\ref{fig:DQDpot}~(b) shows the ground and excited state wavefunctions for $\delta = 0$ in purple.}

\edit{In practice, the potential energy landscape similar to Fig.~\ref{fig:DQDpot} is achieved by combining two materials~\cite{Bjork2002, Nilsson2016, Ranni2024}. The quantum dots are made with a material where the electrons can reside at lower energies as compared to the other material that forms the tunnel barrier. In addition, the voltages applied to gate electrodes change the potential energy $V(x)$. Another usual approach is to use the applied gate voltages to define the tunnel barriers in a DQD structures~\cite{Frey2012, Mi2017, Stockklauser2017}. A sufficiently low temperature is reached for these systems with a dilution refrigerator reaching an electronic temperature of $50$~mK. At this temperature, the thermal energy, $kT = \SI{4}{\micro eV}$, is more than an order of magnitude smaller than a typical level spacing in the range of $0.1 -\SI{1}{meV}$ and charging energy typically of the order of $1$~meV.}

\section{The JC system}

\begin{figure}[t]
    \centering
    \includegraphics[width=0.5\textwidth]{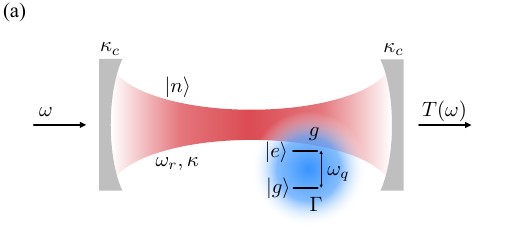}
    \includegraphics[width=0.5\textwidth]{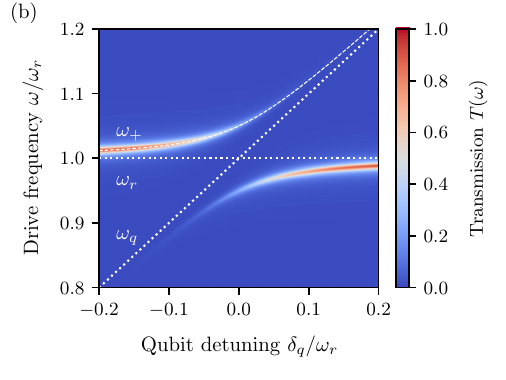}
    \caption{(a) Elements of the considered JC system. A photonic mode (in red) with photon states $\ket{n}$, resonance frequency $\omega_r$, coupling $\kappa_c$ to the input and output port, and total photon loss rate $\kappa$. An electronic TLS with states $\ket{g}$ and $\ket{e}$, energy $\omega_q$, total decoherence rate $\Gamma$, and coupling $g$ to the photonic mode. The transmission $T(\omega)$ is measured as a function of the frequency $\omega$ of the input tone.
    (b) The transmission $T(\omega)$ of Eq.~(\ref{eq:IO1}) as a function of the drive frequency $\omega$ and qubit detuning $\delta_q/\hbar = \omega_q - \omega_r$. The other parameter values are $\kappa_c = 5 \cdot 10^{-3}\, \omega_r$, $\kappa = 2\, \kappa_c$ (i.e. no internal losses), $g = 0.05\, \omega_r$ and $\Gamma = 10^{-2}\, \omega_r$. The response here is typical in the experiments. See e.g. Ref.~\citealp{Delteil2019}.}
    \label{fig:dev}
\end{figure}

Figure~\ref{fig:dev} (a) presents the JC system. 
Here a photonic mode in red with resonance frequency $\omega_r$ is defined in a resonator structure. It couples to an input and output port with couplings $\kappa_c$. The linewidth and therefore also the decoherence rate, $\kappa$ of this mode are set by the sum of these two couplings plus any additional photon losses. The response of the system is determined by measuring the transmission coefficient $T(\omega)$ for a weak input drive signal at frequency $\omega$. With the weak drive, the photon number states $\ket{n}$ are limited to the lowest two states $\ket{0}$ and $\ket{1}$.

The JC Hamiltonian describes the interaction of a TLS with this photon mode~\cite{Jaynes1963, Haroche2006, Blais2004, Childress2004}. The TLS, i.e. a qubit, is shown in blue in Fig.~\ref{fig:dev}~(a). It has an energy difference $\hbar \omega_q$ between the ground state $\ket{g}$ and the excited state $\ket{e}$, and interacts with the photon mode with a coupling rate $g$. The decoherence rate of the TLS is denoted by $\Gamma$. The JC Hamiltonian is then
\begin{equation}
\label{eq:JC}
\begin{array}{ccl}
    
    \displaystyle
    \hat{H}_\mathrm{JC}/\hbar &=& (\hat{H}_r + \hat{H}_\mathrm{DQD} + \hat{H}_\mathrm{int})/\hbar  \vspace{4pt}
    \\
     & = &  \omega_r\, \hat{a}^\dagger \hat{a} +
	 \omega_q \ket{e}\bra{e} + g(\hat{a} \hat{\sigma}^+ + \hat{a}^\dagger \hat{\sigma}^-),
\end{array}
\end{equation}
where $\hat{a}^\dagger$ and $\hat{a}$ are the photon creation and annihilation operators respectively, and $\hat{\sigma}^+ = \ket{e}\bra{g}$ and $\hat{\sigma}^- = \ket{g}\bra{e}$ similarly for the TLS. Here the energy reference is chosen so that the lowest state  $\ket{0} \otimes \ket{g} = \ket{0g}$ is at zero energy. With this choice, the higher energy eigenenergies yield directly the transition energies from this ground state to the corresponding state visible in the transmission spectrum $T(\omega)$. \edit{In order to arrive to the JC Hamiltonian, the coupling rate $g$ is assumed to be much smaller than the photon frequency $\omega_r$ so that the so-called rotating wave approximation (RWA) with only energy conserving terms in the coupling Hamiltonian are relevant.} 

The JC Hamiltonian is obtained for a DQD in the cavity by taking the DQD Hamiltonian of Eq.~(\ref{eq:DQDhamDiag}) and shifting the zero energy reference to $E_-$. This yields immediately the second term in Eq.~(\ref{eq:JC}). The first term $\hat{H}_r$ is obtained by quantizing an electrical LC resonator. This is described thoroughly e.g. in Ref.~\citealp{Girvin2014}, and yields naturally the first term.

\section{Light-matter coupling for a DQD}

The last term $\hat{H}_\mathrm{int}$ in Eq.~(\ref{eq:JC}) describes the interaction between the photons and the TLS. This interaction term for a DQD was first determined in 2004~\cite{Childress2004}. Around the same time, a similar calculation was made for a very similar superconducting charge qubit~\cite{Blais2004}. Here we summarize the general idea, assumptions made, and discuss the experimental realizations briefly. 

\edit{The interaction between the photons in a microwave resonator and the DQD is based on the voltage that a photon in the resonator has. The voltage that a photon generates in the resonator, is highest at the ends of the resonator. Therefore one typically couples one end of the resonator to the DQD to achieve the strongest coupling. The coupling is made by connecting one of the gate electrodes~\cite{Frey2012, Mi2017, Stockklauser2017, Scarlino2022} or the source or drain contact~\cite{Khan2021, Ranni2024} of the DQD to the resonator end. It is important to keep the the wiring much shorter than the wavelength of the photons and the size of the resonator so that the resonance mode remains essentially unchanged. Such a short line makes the voltage $\hat{V}_r$ at the end of the resonator to be present at the gate/contact electrode. This voltage then shifts the detuning $\delta$ of Eq.~(\ref{eq:DQDham}) the same way as the dc voltages do: a positive voltage attracts the electron closer to the gate electrode and the nearby QD. The induced energy shift to $\delta$ is linearly proportional to the applied voltage~\cite{vanderWiel2002}. If we consider that the gate electrode couples to the left QD, then the left dot energy level $\ket{L}$ shifts in energy. This results in the}
interaction Hamiltonian \edit{as}
\begin{equation}
\label{eq:Hint}
\hat{H}_\mathrm{int} = \alpha\, e \hat{V}_r\, \ket{L}\bra{L},
\end{equation}
where $\alpha$ describes \edit{the coupling strength of the gate electrode to the left dot, i.e.} how large energy shift the resonator voltage $\hat{V}_r$ imposes to the left quantum dot energy state $\ket{L}$. The value of $\alpha$ is determined by the capacitances~\cite{Scarlino2022, vanderWiel2002}, and the maximum value, $\alpha = 1$, is achieved if the resonator - left QD capacitance is much larger than any other capacitance to the left QD and the capacitances from the resonator and from the left dot to the right one are small compared to the total capacitance of the right dot. In this case, the state $\ket{L}$ moves by the full amount $e \hat{V}_r$ in energy\edit{.}

In the realized systems, the resonator couples \edit{always} to some degree also to the other QD. \edit{This shifts the right QD energy level $\ket{R}$ and results in a similar coupling term as in Eq.~(\ref{eq:Hint}) with "$L$" replaced by "$R$". As the right dot energy level influences the detuning with the opposite polarity, see Eq.~(\ref{eq:DQDham}), the total coupling is the difference between the coupling strength to the left and right dot. As the DQD is a symmetric structure with respect to the two dots, the coupling to the right and to the left dot must result in the same physical coupling mechanism. Therefore, it suffices to consider only the coupling term of Eq.~(\ref{eq:Hint}) to describe all the physics. Here $\alpha$ is then the overall total coupling strength. It still has the maximum value of $\alpha = 1$ if one of the dots is strongly coupled as described above, and the other dot has a suppressed coupling such that its energy level does not} move at all in response to the resonator voltage $\hat{V}_r$. The constant $\alpha$ is \edit{the so-called lever arm that} describes how strongly a voltage on a gate electrode couples to a given QD.

\edit{Next, we show that Eq.~(\ref{eq:Hint}) results in the coupling term of the JC Hamiltonian in Eq.~(\ref{eq:JC}). We will see which assumptions are required and how the coupling $g$ depends on the DQD and resonator parameters. Towards this end, we need the} resonator voltage $\hat{V}_r$\edit{. As described in detail in Ref.~\citealp{Girvin2014}, it} is obtained from quantizing the LC oscillator as
\begin{equation}
\label{eq:Vres}
\hat{V}_r = \frac{\hbar \omega_r}{e} \sqrt{\frac{\pi Z_r}{R_Q}} \left(a^\dagger + a \right),
\end{equation}
where $Z_r = \sqrt{L/C}$, is the resonator impedance \edit{of a resonator with inductance $L$ and capacitance $C$} and $R_Q = e^2/h$, the resistance quantum. By expressing further the $\ket{L}\bra{L}$ operator of Eq.~(\ref{eq:Hint}) in the $\ket{g}$-$\ket{e}$ -basis yields
\begin{widetext}
\begin{equation}
\label{eq:HintFull}
\hat{H}_\mathrm{int} = \hbar \omega_r\, \alpha\, \sqrt{\frac{\pi Z_r}{R_Q}} \left(a^\dagger + a \right)
\left(\frac{\delta/2}{\hbar\omega_r}\left(\ket{e}\bra{e} - \ket{g}\bra{g}\right) + 
\frac{t}{\hbar \omega_r} \left(\ket{e}\bra{g} + \ket{g}\bra{e}\right) + 
\frac{1}{2}  \right).
\end{equation}
\end{widetext}
\edit{The JC Hamiltonian of Eq.~(\ref{eq:JC}) is obtained from this result by making the rotating wave approximation (RWA), i.e. keeping only energy conserving terms.} The resulting light-matter coupling constant $g$ in Eq.~(\ref{eq:JC}) is then
\begin{equation}
\label{eq:g}
g = \hbar \omega_r\, \alpha\, \sqrt{\frac{\pi Z_r}{R_Q}} \frac{t}{\hbar \omega_r}. 
\end{equation}
Here $Z_r$ is the impedance of the LC circuit. The result of Ref.~\citealp{Childress2004} deviates from this by a small factor of the order of unity as it uses the characteristic impedance of a transmission line resonator instead of the impedance of the resonance mode~\cite{Goppl2008}.

\edit{The RWA and t}he JC Hamiltonian of Eq.~(\ref{eq:JC}) with the light-matter coupling constant of Eq.~(\ref{eq:g}) apply if the coupling constant is small $g \ll \omega_q ,\omega_r$. In this case, the interaction is strong only around the resonant condition $\delta_q/\hbar \equiv  \omega_q - \omega_r = 0$. Note that the detuning $\delta_q$ here is the energy difference between the total energy $\hbar \omega_q$ of the \edit{TLS} and the photon energy, not the detuning $\delta$ of the localized DQD states $\ket{L}$ and $\ket{R}$. At large detuning $\delta_q$, the interaction between the TLS and the resonator is suppressed for $g \ll \omega_q ,\omega_r$. For larger coupling constants $g \gtrsim 0.1 \omega_r$, one reaches the so-called ultra-strong coupling regime~\cite{FriskKockum2019}. In this case, the \edit{RWA is not valid and} non-energy conserving terms become relevant. Here it is noteworthy that one needs to keep the full form of Eq.~(\ref{eq:HintFull}) for this regime. This gives rise to additional coupling mechanisms. In addition to the so-called transversal coupling of the form of $\left(a^\dagger + a \right)\left(\ket{e}\bra{g} + \ket{g}\bra{e}\right)$ that contains the energy conserving terms, one also has the so-called longitudinal coupling of the form $\left(a^\dagger + a \right)\left(\ket{e}\bra{e} - \ket{g}\bra{g}\right)$~\cite{Childress2004}.

In experiments, ordinary $50~\Omega$ transmission line resonators with resonance frequencies in the most typical measurement band of $\omega_r = 4 ... 8$~GHz have $g/2\pi \approx 10$'s~MHz~\cite{Frey2012, Mi2017}. For these systems, the RWA is satisfied. More recently, high impedance resonators have been employed. These resonators have a higher characteristic impedance of the order of $Z_r = 1~\mathrm{k\Omega}$ increasing the coupling to $g/2\pi \approx 100$'s~MHz~\cite{Stockklauser2017, Scarlino2022,Ranni2023}. In this case, $g \sim 0.1\, \omega_r$, i.e. these systems start to approach the ultra-strong coupling regime where corrections to the RWA start to become relevant.

\section{Input-output theory}
To determine the transmission $T(\omega)$ of the JC system, the so-called input-output theory~\cite{Gardiner1984} is typically used. See e.g. Refs.~\citealp{Girvin2014, Blais2021} for a detailed derivation. The input-output theory describes the signal lines connecting to the system with with the couplings $\kappa_c$ of Fig.~\ref{fig:dev} and calculates the outgoing signals for given input signals. We denote these as $\hat{b}_{\mathrm{in},k}(t)$ and $\hat{b}_{\mathrm{out},k}(t)$ for the $k$-th line. \edit{The input-output formalism assumes that the coupling rates $\kappa_c$ are small compared to the photon energy $\omega_r$. Furthermore, all the losses in the resonator including the coupling to the input/output line are assumed to be frequency independent. This assumption is equivalent to the Markov approximation that the input/output lines do not store any information that would be fed back to the system at a later time. With these assumptions,} the theory yields
\begin{equation}
\label{eq:IOgen1}
\hat{b}_{\mathrm{out},k}(t) = \hat{b}_{\mathrm{in},k}(t) + \sqrt{\kappa_c} \hat{a},
\end{equation}
which can be interpreted as a continuity condition that at the coupling port, the outgoing signal consists of the direct reflection of the ingoing signals plus the signal emitted \edit{from} the system with a rate $\kappa_c$. In addition, the input output theory yields \edit{the Heisenberg} equation of motion for $\hat{a}$, namely
\begin{equation}
\label{eq:IOgen2}
\dot{\hat{a}} = \frac{i}{\hbar} \left[\hat{H}_S, \hat{a} \right] - \frac{\kappa}{2} \hat{a} - \sqrt{\kappa_c}\ \sum_k\hat{b}_{\mathrm{in},k}(t).
\end{equation}
Here the first term on the right hand side describes the unitary evolution with the system Hamiltonian $\hat{H}_S$. For our case $\hat{H}_S = \hat{H}_\mathrm{JC}$. The second term describes damping with of the optical mode with the total rate $\kappa$, and the last one the drive from the input signal. \edit{Here we have assumed the signal lines to have the same coupling $\kappa_c$.} More generally these can be taken with a different value for each line in each term. For the transmission response, we consider next the case where the system is driven only from line $k=1$ with $\hat{b}_{\mathrm{in},1}(t)$ and the transmission is measured from line $k=2$ with the out-going signal $\hat{b}_{\mathrm{out},2}(t)$ and no input signal, i.e. $\hat{b}_{\mathrm{in},2}(t) = 0$.
Inserting Eq.~(\ref{eq:JC}) to Eq.~(\ref{eq:IOgen2}) yields then
\begin{equation}
\label{eq:IOinterm1}
\dot{\hat{a}} = -i\omega_r \hat{a} - ig \hat{\sigma}
- \frac{\kappa}{2} \hat{a} - \sqrt{\kappa_c}\ \hat{b}_{\mathrm{in},1}(t).
\end{equation}
We see that the time evolution of $\hat{a}$ depends on the DQD via the operator $\hat{\sigma} = \ket{g}\bra{e}$. We therefore need also the time evolution of $\hat{\sigma}$. We \edit{obtain} this with the Lindblad equation in the Heisenberg picture as
\begin{equation}
\label{eq:IOsigma1}
\dot{\hat{\sigma}} = \frac{i}{\hbar} \left[\hat{H}_S, \hat{\sigma} \right] - \frac{\Gamma_0}{2} \hat{\sigma},
\end{equation}
where $\Gamma_0$ is the total dephasing rate of the DQD. Again, inserting Eq.~(\ref{eq:JC}) yields
\begin{equation}
\label{eq:IOsigma2}
\dot{\hat{\sigma}} = -i\omega_q\hat{\sigma} + ig\hat{a}\hat{\sigma}_z
- \frac{\Gamma_0}{2} \hat{\sigma},
\end{equation}
where $\sigma_z = \ket{e}\bra{e} - \ket{g}\bra{g}$. Equations~(\ref{eq:IOinterm1}) and (\ref{eq:IOsigma2}) constitute a set of linear differential equations apart from the term $ig\hat{a}\hat{\sigma}_z$. This term is linearized \edit{with the approximation $\hat{\sigma}_z = \left<\hat{\sigma}_z\right>$, and assumption that $\left<\hat{\sigma}_z\right>$ is constant during the considered time interval. In other words, we factorize the $ig\hat{a}\hat{\sigma}_z$ term with a mean field approximation. It assumes that the qubit average value $\left<\hat{\sigma}_z\right>$ suffices to describe the light matter coupling in Eq.~(\ref{eq:IOsigma2}).} This \edit{assumption holds} particularly well if the input drive is weak so that the DQD stays most of the time in its ground state $\ket{g}$ with $\hat{\sigma}_z = -1$. \edit{It is also applicable if other external conditions such as the temperature of the system, or the preparation of the TLS state determine $\left<\hat{\sigma}_z\right>$ and it does not change significantly during the considered time interval. A simple example is the relaxation of the TLS. The response needs to be considered at a time interval much shorter than the relaxation time. Otherwise $\left<\hat{\sigma}_z\right>$ evolves over time and this needs to be taken into account. 
In addition, a strong drive may induce correlations into the system. In this case, the approximation $\hat{\sigma}_z = \left<\hat{\sigma}_z\right>$ is not sufficient to describe the response, and Eq.~(\ref{eq:IOsigma2}) cannot be linearized. One needs to calculate the correlations arising from the $\hat{a}\hat{\sigma}_z$ term explicitly with numerical calculations. Numerical calculations are often required also for checking the validity of the made approximations for the analytic results.

With the above mean field approximation,} we can Fourier transform the equations and obtain
\begin{equation}
\label{eq:IOinterm2}
\left\{
\begin{array}{ccl}
      -i\omega\hat{a}(\omega) &=& \displaystyle -i\omega_r \hat{a}(\omega) - ig \hat{\sigma}(\omega)
- \frac{\kappa}{2} \hat{a}(\omega) - \sqrt{\kappa_c}\ \hat{b}_{\mathrm{in},1}(\omega) \\
     -i\omega\hat{\sigma}(\omega) &=& \displaystyle -i\omega_q\hat{\sigma}(\omega) + ig\hat{a}(\omega)\left<{\sigma}_z\right>
- \frac{\Gamma_0}{2} \hat{\sigma}(\omega).
\end{array}
\right.
\end{equation}
Now we can finally take the expectation values of these equations and calculate the transmission coefficient as
\begin{equation}
\label{eq:IO1}
T(\omega) \equiv \left| \frac{\left< \hat{b}_{\mathrm{out},2}(\omega) \right>}{\left< \hat{b}_{\mathrm{in},1}(\omega) \right>} \right| = \left| \kappa_c A\left(\omega\right) \right|^2,
\end{equation}
where
\begin{equation}
\label{eq:IO2}
A(\omega) = \frac{\Gamma/2 - i\left(\omega-\omega_q\right)}
{\left[ \kappa/2 - i\left(\omega-\omega_r\right) \right] \left[ \Gamma/2 - i\left(\omega-\omega_q\right) \right] - g^2 \left<\sigma_z\right> }.
\end{equation}
For the weak drive and low temperatures, the qubit stays majority of the time in the ground state and $\left<\sigma_z\right> \approx -1$. We consider \edit{predominantly} this limit in the following. \edit{If the temperature of the system is high, $T \gtrsim \hbar \omega_q/k$, or the system driven stronger, the TLS will be excited. For high temperature or strong drive, the ground and excited states tend to be equally populated. In this case $\left<\sigma_z\right> \rightarrow 0$, and the interaction becomes suppressed~\cite{Abdumalikov2010, Astafiev2010}. The suppression is seen in Eq.~(\ref{eq:IO2}) as it results in the bare resonator response, $A(\omega) = 1/[\kappa/2 - i\left(\omega-\omega_r\right)]$, in this limit. Furthermore, strong drive may induce nonlinear effects and correlations between the TLS and the photonic mode, which may lead to a more complex response~\cite{Baur2009, Kirchmair2013, Andersson2025}. In this case, the photonic mode description presented in this review is not anymore applicable.}

Figure~\ref{fig:dev} (b) shows the resulting transmission for a typical case in the strong-coupling regime, $g > \Gamma, \kappa$ as a function of input drive frequency $\omega$ and \edit{TLS/}qubit detuning $\delta_q = \omega_q - \omega_r$. At large detuning $|\delta_q| \gg g$, the eigenstates of the system are close to the bare states $\ket{n} \otimes \ket{q} = \ket{nq}$ with the photon number $n = 0, 1, ...$, and the qubit state $q = g, e$, and only the $\ket{0} \leftrightarrow \ket{1}$ transition is visible in the response at $\omega \approx \omega_r$. For small detuning, the states hybridize and show the Rabi-split avoided crossing. As well known in the field, the lowest eigenenergies are obtained by  diagonalizing the JC Hamiltonian in the $\ket{0e}$ - $\ket{1g}$ basis resulting in
\begin{equation}
\label{eq:omegapm}
\omega_\pm = (\omega_q + \omega_r)/2 \pm \Omega/2,
\end{equation}
where $\Omega = \sqrt{\delta_q^2 + 4 g^2}$. As seen in the figure, the transmission has strong resonant response at these energies. The corresponding eigenstates are
\begin{equation}
\label{eq:psipm}
\ket{\psi_\pm} = c_\pm \ket{0e} \pm c_\mp \ket{1g},
\end{equation}
where $c_\pm = \sqrt{(1 \pm \delta_q/\Omega)/2}$. These are again the Hopfield coefficients~\cite{Hopfield1958, Kavokin2015, Kavokin2017, Ferreira2022} that we encoutered already in the section describing the DQD hybridization with the equivalent mixing angle expressions $c_+ = \cos(\phi/2)$ and $c_- = \sin(\phi/2)$ and $\phi = \arctan(2g/\delta_q)$, that are more frequently used expression in the microwave domain~\cite{Haroche2006, Blais2004, Childress2004}. Note also that here we consider a system with only one TLS such that inhomogeneous broadening is not relevant. The solid-state devices with microwave photons have so large dipole coupling that already a single TLS achieves the strong coupling regime~\cite{Wallraff2004, Mi2017, Stockklauser2017, Burkard2020, Ranni2023, Ungerer2024}.

\section{Modes in the strong-coupling case}
\label{sect:strongCoupl}

\edit{Next we consider the two states of Eq.~(\ref{eq:psipm}) as separate modes and determine their resonance frequency, linewidth and coupling strength present in the transmission response, all at the same time.} To proceed, we \edit{first} limit our consideration to the strong-coupling case for which \edit{$g > \Gamma, \kappa$}. In this case, the eigenenergies are well separated from each other, and their response around the frequencies $\omega_\pm$ is far away from $\omega_q$. In the following, we consider the response $T(\omega)$ nearby these modes allowing us to approximate $\Gamma/2 - i(\omega-\omega_q) = i(\omega-\omega_q) = i(\omega_\pm-\omega_q)$ in the nominator of Eq.~(\ref{eq:IO2}). With this, we obtain
\begin{widetext}
\begin{equation}
        A\left(\omega\right) = \frac{\omega_\pm - \omega_q}
{\kappa/2\left(\omega-\omega_q\right) + \Gamma/2\left(\omega-\omega_r\right)  - i\left(\omega-\omega_+\right) \left(\omega-\omega_-\right) }
= \frac{\omega_\pm - \omega_q}
{\kappa/2\left(\omega_\pm-\omega_q\right) + \Gamma/2\left(\omega_\pm-\omega_r\right)  \mp i \Omega \left(\omega-\omega_\pm\right) },
    \label{eq:Ainterm}
\end{equation}
\end{widetext}
where the upper choice is for the $\omega_+$ mode around $\omega = \omega_+$ and the lower one for the $\omega_-$ mode around $\omega = \omega_-$. Here we have neglected the term $\kappa\Gamma/4 \ll g^2$ which gives rise to a small shift to the resonance frequency in form of a renormalization of $g$. Next we note that the energy differences are connected to the hybridization of the photonic and electronic states via the wave function coefficients $c_\pm$ as $\omega_\pm - \omega_r = \pm\Omega\, \left| c_\pm \right|^2$ and $\omega_\pm - \omega_q = \pm\Omega\, \left| c_\mp \right|^2$. Therefore, we obtain finally with Eqs.~(\ref{eq:IO1}) and (\ref{eq:Ainterm}) the transmission coefficient of the system as
\begin{equation}
\label{eq:Lorentzian}
T\left(\omega\right) = \frac{\kappa_{c\pm}^2}{\left(\kappa_\pm/2\right)^2 + \left(\omega - \omega_\pm\right)^2},
\end{equation}
where $\kappa_{c\pm} = \kappa_c \left| c_\mp \right|^2$ \edit{is the coupling of the mode} and $\kappa_\pm = \kappa \left| c_\mp \right|^2 + \Gamma \left| c_\pm \right|^2$ \edit{the sum of all the losses of the mode that sets the linewidth of the mode}. Equation~(\ref{eq:Lorentzian}) is a Lorentzian, and together with the analytical expressions of $\kappa_{c\pm}$ and $\kappa_\pm$, \edit{is central for the mode description of the JC system}. It has identical structure as the transmission of the bare resonator has with $T\left(\omega\right) = \kappa_c^2/( (\kappa/2)^2 + \left(\omega - \omega_r\right)^2 )$, see e.g. Refs.~\citealp{Goppl2008, Ranni2023}. In Eq.~(\ref{eq:Lorentzian}) each of the two states $\ket{\psi_\pm}$ of Eq.~(\ref{eq:psipm}) gives rise to a Lorentzian response with the $\ket{0g} \leftrightarrow \ket{\psi_\pm}$ transition at resonance frequency $\omega_\pm$. The \edit{couplings} of these modes are given by the original photonic coupling $\kappa_c$ scaled with the wavefunction weight $\left| c_\mp \right|^2$ of the bare photonic excitation component $\ket{1g}$ of Eq.~(\ref{eq:psipm}). Similarly, the linewidth of the mode $\kappa_\pm$, i.e. total decoherence rate of the mode, is a sum of the photonic contribution with the rate $\kappa$ and the same photonic weight, and the corresponding electronic contribution with the rate $\Gamma$ and the corresponding wavefunction weight $\left| c_\pm \right|^2$ of the bare electronic excitation component $\ket{0e}$. With these results, we get for example in the resonant case $\delta_q = 0$\edit{,} the well-known result $\kappa_\pm = (\kappa + \Gamma)/2$, i.e. that the decoherence rate for the fully hybridized states is the average of the electronic and photonic decoherence rate~\cite{Blais2021, Ranni2024}. The couplings of the modes derived here \edit{allow} to obtain further insights to the interaction dynamics, and \edit{connect} the \edit{peak transmission} of the hybridized states and the coherence of the electronic system together as we discuss next.

\begin{figure}[t]
    \centering
    \includegraphics[width=0.5\textwidth]{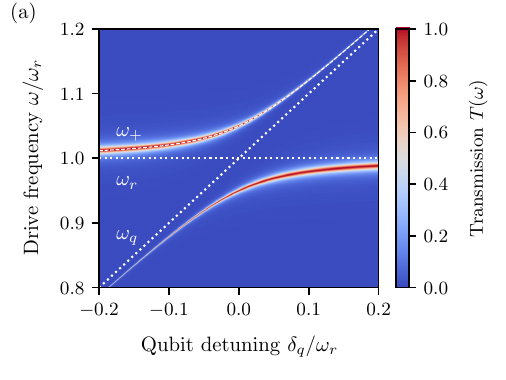}
    \includegraphics[width=0.5\textwidth]{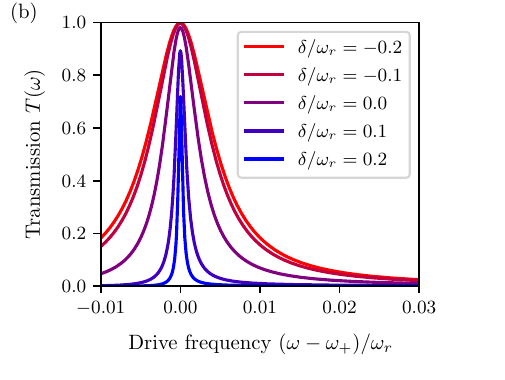}
    \caption{(a) The transmission $T(\omega)$ of Eq.~(\ref{eq:IO1}) for a high-coherence TLS with $\kappa_c = 5 \cdot 10^{-3}\, \omega_r$, $\kappa = 2\, \kappa_c$, $g = 0.05\, \omega_r$ and $\Gamma = 10^{-4}\, \omega_r$.
    (b) The transmission $T(\omega)$ plotted as a function of drive frequency $\omega$ around the mode at frequency $\omega_+$. The red data is at $\delta_q < 0$ where the mode is predominantly photonic and the blue one at $\delta_q > 0$ where the mode is predominantly electronic. For \edit{the electronic side}, the transmission stays close to unity while the linewidth decreases significantly.}
    \label{fig:RespCoh}
\end{figure}

\begin{figure}[t]
    \centering
    \includegraphics[width=0.5\textwidth]{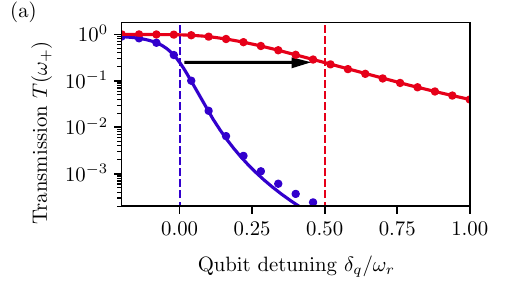}
    \includegraphics[width=0.5\textwidth]{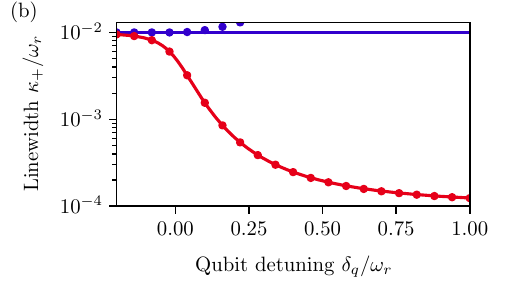}
    \caption{The \edit{peak} transmission $T(\omega_+)$, panel (a), and linewidth $\kappa_+$, panel (b), for the mode $\omega_+$ for different detunings. The blue data is for the case in Fig.~\ref{fig:dev} and the red data for the Fig.~\ref{fig:RespCoh} case. The dots are determined from the full response of Eq.~(\ref{eq:IO1}), while the lines show the result of Eq.~(\ref{eq:Lorentzian}). The results for the mode $\omega_-$ are the same with mirroring $\delta_q \mapsto -\delta_q$. The dashed lines denote the condition $\kappa \left| c_\mp \right|^2 = \Gamma \left| c_\pm \right|^2$ where the linewidth has equal contributions from the photonic and electronic part. At this condition, the \edit{peak} transmission is reduced by a factor of $4$.} 
    \label{fig:TandKappa}
\end{figure}

In Fig.~\ref{fig:dev} (b), the photonic and electronic coherence is chosen equal, $\kappa = \Gamma$, such that the decoherence rate, i.e. the effective internal losses, of the hybridized states remains constant $\kappa_\pm = \kappa$. The \edit{peak transmission} of the bare photonic mode then decreases as the state hybridizes and turns more into the electronic one. The reduction of the \edit{transmission} is because the input/output coupling reduces \edit{for increasing contribution from} the electronic part. The bare photonic state has the highest \edit{peak transmission} of $T(\omega_\pm) = 4\kappa_c^2/\kappa^2$. As the states hybridize and obtain larger and larger electronic fraction, the transmission reduces as $T(\omega_\pm) = 4\kappa_c^2 \left| c_\mp \right|^4/\kappa^2$.

Figure~\ref{fig:RespCoh} \edit{represents} another interesting case. Here all the other parameter values are the same \edit{as in Fig.~\ref{fig:dev}~(b)} except the decoherence rate $\Gamma$ of electronic state which is reduced to have $\Gamma \ll \kappa, \kappa_c$, such that the electronic system is more coherent than the photonic one. In this case, the non-hybridized photonic states have still the \edit{peak transmission} of $T(\omega_\pm) = 4\kappa_c^2/\kappa^2$. However, as these states hybridize with the electronic one, the transmission stays high, as see in Fig.~\ref{fig:RespCoh} (a) for the states close to $\omega_q$, and in the corresponding line cuts of Fig.~\ref{fig:RespCoh} (b). At the same time, the Lorentzian becomes sharper with a smaller linewidth. These effects can be again understood with the couplings of Eq.~(\ref{eq:Lorentzian}). For small $\Gamma$, both the $\kappa_{c\pm}$ and $\kappa_\pm$ scale proportional to $\left| c_\mp \right|^2$. Therefore both the coupling and the total decoherence rate reduce in concert\edit{,} leading to higher coherence and remaining high transmission \edit{as} the state becoming more electronic.

Figure~\ref{fig:TandKappa} summarizes the above findings by presenting the maximum transmission $T(\omega_+)$ and the linewidth $\kappa_+$ for the two cases. The dots show the transmission and linewidth extracted directly from Eq.~(\ref{eq:IO1}), and the lines the corresponding values based on the Lorentzian of Eq.~(\ref{eq:Lorentzian}). Blue data is for the $\Gamma = \kappa$ case and red one for the $\Gamma \ll \kappa,  \kappa_c$. We see that the Lorentzian approximation catches the main features of the data: The \edit{peak transmission} $T(\omega_+)$ starts at a close-to-unity value and reduces at around $\delta_q \approx 0$ for the blue data of the $\Gamma = \kappa$ case as the bare resonator state hybridizes significantly with the electronic contribution. At the same time, the total linewidth stays approximately constant. At large positive $\delta_q$, the state becomes predominantly electronic like and the transmission is suppressed. In this case, the precise response deviates from the prediction of Eq.~(\ref{eq:Lorentzian}) since the tails of the much stronger $\ket{0g} \leftrightarrow \ket{\psi_-}$ transition contribute significantly at around $\omega_+$. This leads to larger transmission $T(\omega_+)$ and increased estimated linewidth as seen in Figs.~\ref{fig:TandKappa}~(a) and (b) respectively. This regime is, however, not particularly relevant as the response is already suppressed and typically not resolvable in the experiments any more~\cite{Wallraff2004, Stockklauser2017, Ranni2024}.

The red points and lines of Fig.~\ref{fig:TandKappa} repeat the summary for the coherent case of Fig.~\ref{fig:RespCoh}. Again, the solid lines of Eq.~(\ref{eq:Lorentzian}) predict the response accurately. Now, instead of the transmission $T(\omega_+)$ decreasing at around $\delta_q \approx 0$, rather the linewidth starts decreasing and the resonance sharpens here as the hybridization with the electronic part sets in. The high transmission coefficient and sharpening of the response continue until $\kappa_+ \approx \Gamma$. After this point, the linewidth saturates towards a constant value of $\Gamma$ and the transmission starts reducing analogous to the less coherent case of the blue data. The onset point for the reduction of $T(\omega_+)$ shifts from the $\delta_q \approx 0$ point to the condition $\kappa \left| c_\mp \right|^2 = \Gamma \left| c_\pm \right|^2$, i.e. to the point where the decoherence of the photonic part has equal contribution to $\kappa_\pm$ as the electronic contribution. Under this condition, the ratio $\kappa_\pm/\kappa_{c\pm}$ doubles
and leads to a factor of $4$ reduction in $T(\omega_\pm)$, see Eq.~(\ref{eq:Lorentzian}). With the assumption $\Gamma \ll \kappa$, this takes place at $\delta_q = \sqrt{\kappa/\Gamma}\, g$. For the presented data, we obtain $\delta_q/\omega_r = 0.5$ as indicated with the black arrow in Fig.~\ref{fig:TandKappa}~(a).

\section{Dispersive shift}
Dispersive shift is one of the most commonly used \edit{method to measure} the qubit \edit{state}~\cite{Blais2004}. It is based on a frequency shift \edit{imposed by the qubit} to the resonator\edit{.} The frequency shift is $\Delta \omega_r = \pm g^2/\delta_q$, where the plus sign applies for the excited state and the minus sign for the ground state. Here the detuning $\delta_q = \omega_q-\omega_r$, is assumed to be large: $\delta_q \gg g$. \edit{As the frequency shift is in opposite directions for the ground and excited state, probing the resonance frequency allows for inferring the qubit state.}  

A usual way to derive this shift is to make a unitary transformation to the \edit{JC} Hamiltonian~\cite{Boissonneault2009}. Here we consider the response given by Eqs.~(\ref{eq:IO1}) - (\ref{eq:IO2}), and show that the dispersive shift arises as a limiting case. For that, \edit{we lift the assumption $\left<\sigma_z\right> = -1$, that the qubit is in the ground state, and assume instead a fixed constant value $\left<\sigma_z\right>$ that does not change during the readout. W}e consider \edit{the mode that is nearly the bare resonator mode with its resonance frequency} close to \edit{$\omega_r$. For the response close to this resonance, we have} $\omega \approx \omega_r$ \edit{and can therefore approximate} $\omega-\omega_q \approx \omega_r-\omega_q = -\delta_q$ in Eq.~(\ref{eq:IO2}). \edit{Assuming further a large detuning}, $\left|\delta_q\right| \gg \Gamma/2$, yield\edit{s} a Lorentzian lineshape
\begin{equation}
\label{eq:Adisp}
A(\omega) = \frac{1}
{\kappa/2 - i\left(\omega-\omega_r - g^2 \left<\sigma_z\right>/\delta_q \right)},
\end{equation}
which is the same as the bare resonator response ($g=0$) with the additional frequency shift $\Delta \omega_r = g^2 \left<\sigma_z\right>/\delta_q$. \edit{This matches the above frequency shift $\Delta \omega_r = \pm g^2/\delta_q$ as $\left<\sigma_z\right> = 1$ for the excited state and $\left<\sigma_z\right> = -1$ for the ground state. Note also that the linewidth and couplings of Eq.~(\ref{eq:Adisp}) are the same as for the bare resonator. This is because the electronic state does not hybridize significantly with the considered photonic mode and hence the linewidth and couplings remain unchanged. It is also worth noting that Equation~(\ref{eq:Adisp}) coincides with Eq.~(\ref{eq:Lorentzian}) when considering the mode closer to $\omega_r$ and taking the limit $\delta_q \gg g$.}

\edit{The above assumption that the expectation value $\left<\sigma_z\right>$ does not change during the readout is satisfied if the readout time is shorter than the qubit relaxation time, and the resonator is driven only weakly.} The unitary transformation \edit{method}~\cite{Blais2004, Boissonneault2009}, shows that the dispersive shift is valid up to a photon number $\left< n\right> = \delta_q^2/4g^2$. It also keeps track explicitly the operators in the calculation, which is important e.g. for showing that this readout method achieves the quantum limit \edit{as a quantum non-demolition measurement~\cite{Clerk2010}}, and for determining how reading out the resonator impacts the qubit. The \edit{above} considerations based on Eqs.~(\ref{eq:IO1}) - (\ref{eq:IO2})  present another aspect that is relevant: the dephasing rate of the qubit needs to be small $\Gamma \ll 2 \left|\delta_q\right|$. If this is not valid, the response will show additional losses \edit{and the dispersive frequency shift will be suppressed. This regime is considered below in the section describing the weak-coupling limit, see Eqs.~(\ref{eq:loss}) and (\ref{eq:shift}).}

For completeness, we summarize the main idea and the resulting Hamiltonian with the unitary transformation method of Refs.~\citealp{Blais2004, Boissonneault2009}. The Jaynes-Cummings Hamiltonian in the dispersive limit is obtained \edit{by} making a unitary transformation $U = \exp\left(\lambda X_- \right)$ with $X_- = a^\dagger \sigma^- - a \sigma^+$ to Eq.~(\ref{eq:JC}). Choosing $\lambda = g/\delta_q$, and keeping only terms in the leading order of $\lambda$ yields the Hamiltonian in the dispersive limit
\begin{equation}
\label{eq:JCdisp}
\hat{H} = \left(\omega_r\ + \frac{g^2}{\delta_q} \edit{\hat{\sigma}_z} \right)\hat{a}^\dagger \hat{a} +
	 \left( \omega_q + \frac{g^2}{\delta_q}\right) \ket{e}\bra{e}.
\end{equation}
The first term contains the dispersive shift. It can be also interpreted as a Stark shift such that the photon number $a^\dagger a$ shifts the qubit energy (via \edit{$\hat{\sigma}_z$}). The term $g^2/\delta_q$ in the second term shifts the qubit frequency $\omega_q$ even in the absence of photons $\left< a^\dagger a \right> = 0$, i.e. just the presence of the resonator changes the frequency. This is know as the Lamb shift.

The approximate limit to the maximum photon number is obtained by considering that the exponent in the unitary transformation needs to be small. As $\left< X_- \right> \sim \left< a \right> \sim \sqrt{\edit{\left< n \right>}}$, this condition is approximately $\lambda \sqrt{\edit{\left< n \right>}} \ll 1$, which yields $\edit{\left< n \right>} \ll \delta_q^2/g^2$.

\section{Resonator mode in the weak-coupling limit}

\begin{figure}[t]
    \centering
    \includegraphics[width=0.5\textwidth]{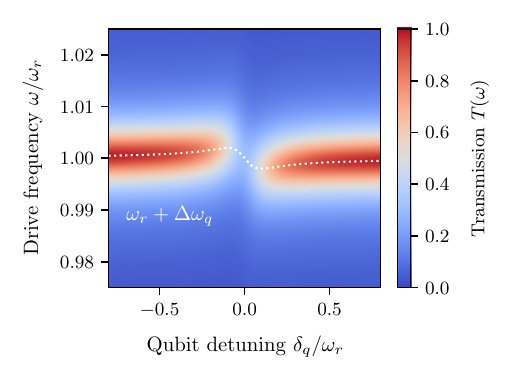}
    \caption{The transmission $T(\omega)$ of Eq.~(\ref{eq:IO1}) in the weak coupling limit with $\kappa_c = 5 \cdot 10^{-3}\, \omega_r$, $\kappa = 2\, \kappa_c$, $g = 0.02\, \omega_r$ and $\Gamma = 0.2\, \omega_r$. The white dashed line indicates the shifted resonance frequency $\omega_r + \Delta \omega_q$ of Eq.~(\ref{eq:shift}).} 
    \label{fig:weakCoupl}
\end{figure}

When the dephasing rate $\Gamma$ exceeds the coupling $g$, the resonator response consists of only one mode near the bare resonator frequency~\cite{Frey2012} as shown in Fig.~\ref{fig:weakCoupl}. \edit{For the response close to this mode,} we can rewrite Eq.~(\ref{eq:IO2}) as
\begin{equation}
\label{eq:Awk}
    A(\omega) = \frac{1}
{\left[ \kappa/2 - i\left(\omega-\omega_r\right) \right] + g \chi },
\end{equation}
where $\chi = g\left< \sigma_z \right>/ \left[ \Gamma/2 - i\left(\omega-\omega_q\right) \right]$ is the electronic susceptibility of the TLS~\cite{Burkard2020}. For the considered case, $\Gamma > g$ the susceptibility is small, i.e. $\left| \chi \right|/2 < 1$. We can then proceed by considering the response close to the bare resonator frequency, $\omega \approx \omega_r$, and determine the real and imaginary part of the $\chi$ \edit{simultaneously}. These are the added dissipation $\kappa_q$ and dispersive shift $\Delta \omega_q$ \edit{respectively} that read as
\begin{equation}
    \label{eq:loss}
    \kappa_q = \frac{4 g^2 \left< \sigma_z \right>}{\Gamma} \frac{1}{1+(2\delta_q/\Gamma)^2},
\end{equation}
and
\begin{equation}
     \label{eq:shift}
    \Delta \omega_q = -\frac{2 g^2 \left< \sigma_z \right>}{\Gamma} \frac{2\delta_q/\Gamma}{1+(2\delta_q/\Gamma)^2}.
\end{equation}
The response function $A(\omega)$ reads then as
\begin{equation}
    \label{eq:Aonemode}
    A(\omega) = \frac{1}
{(\kappa+\kappa_q)/2 - i\left(\omega-\omega_r-\Delta \omega_q \right)}.
\end{equation}
This equation is the bare resonator Lorentzian response with added loss $\kappa_q$ and resonance frequency shift $\Delta \omega_q$. It reproduces the full response of Fig.~\ref{fig:weakCoupl} with vanishingly small differences ($\left| \Delta T(\omega) \right| < 0.02$). We can further see from Eq.~(\ref{eq:loss}) that the maximum loss $\kappa_q = 4g^2/\Gamma$ is obtained at zero detuning $\delta_q = 0$ in line with the double quantum loss used in Ref.~\citealp{Khan2021}. At finite detuning, we have the additional coefficient $1/\left(1+(2\delta_q/\Gamma)^2\right)$. This prefactor is that of the Purcell effect, i.e. that the losses increase when an additional lossy resonance mode is brought close in detuning~\cite{Houch2008}. Note that here we consider the losses added by the TLS to the resonator, which is the opposite to Ref.~\citealp{Houch2008} where the losses of the resonator influencing the qubit coherence was considered. The effect works the same way in both directions. The dispersive shift of Eq.~(\ref{eq:shift}) has a maximal value of $\Delta \omega_q = \pm g^2/\Gamma$ for $\delta_q = \pm \Gamma/2$. For $\delta_q = 0$, the dispersive shift vanishes.

\section{Quantum capacitance}
\begin{figure}[t]
    \centering
    \includegraphics[width=0.3\textwidth]{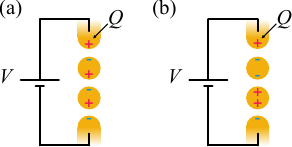}
    \caption{(a) Charge distributions for the geometric capacitance. The two quantum dots illustrated with the yellow circles have overall no net charge added to them with the applied voltage $V$. The charges within the dots are however redistributed as illustrated with the plus and minus signs. As a result, a positive charge $Q$ is accumulated to the positive terminal and corresponding negative charge to the other side. (b) Quantum capacitance contribution. The additional quantum capacitance arises because of net charge changes. Here this is illustrated by an electron tunneling from the lower dot to the higher one such that the higher dot has a negative net charge and the lower one a positive one.} 
    \label{fig:caps}
\end{figure}

Quantum capacitance is \edit{an additional capacitive contribution} increas\edit{ing} the total capacitance to a value larger than the usual \edit{geometric} capacitance~\cite{Luryi1988, Vigneau2023}. It arises as the electron occupancy \edit{changes and results in an additional change in the charge distribution}. For the DQD case considered here, this takes place via the interdot tunneling process. The change in occupancy is induced with an applied voltage. The difference between the geometric and quantum capacitance is illustrated in Fig.~\ref{fig:caps} with the DQD structure. In the geometric capacitance, the the total charge of each island stays fixed and the capacitance arises from the charges redistributing within the islands as illustrated in panel (a). For the quantum capacitance case of panel (b), the occupancy of the islands changes and yields an additional capacitance contribution. This is illustrated with an electron that tunneled from the lower island to the upper one. In reality, the average occupancy of the quantum dots changes gradually with gate voltage with the electrons tunneling many times back and forth \edit{stochastically}, or as a result of the charge distribution changes of the hybridized electronic states. The latter picture applies to the JC system of this review with a coherent electronic system. This is in strong contrast to the first stochastic case that is more usually considered at low frequencies and for a weak drive~\cite{Luryi1988, Vigneau2023}.

One of the most common ways to probe \edit{the quantum capacitance} is to connect the \edit{quantum} device to a resonator and determine the frequency or phase shift that the quantum capacitance imposes to the resonator~\cite{Petersson2010, Gonzalez-Zalba2016, Chatterjee2018}. Here we consider next the frequency shift of Eq.~(\ref{eq:shift}) of the JC system, determine the corresponding quantum capacitance, and compare to the above results in the 'classical' limit. The JC approach \edit{illustrates} how the resonator mode with a finite frequency, as well as the dephasing with rate $\Gamma$ influences the quantum capacitance of a TLS.

The frequency shift $\Delta \omega_\edit{q}$ of Eq.~(\ref{eq:shift}) can be assimilated to a corresponding quantum capacitance $C_q$ by considering that a  capacitance change $\Delta C$ in the resonator capacitance $C$ induces the frequency shift. With the equation $\omega_r = 1/\sqrt{LC}$, the resulting frequency shift is then
\begin{equation}
    \Delta \omega_\edit{q} = -\frac{\Delta C}{2 \sqrt{LC^3}} = -\frac{Z_r \omega_r^2}{2} \Delta C,
\end{equation}
with the characteristic impedance $Z_r = \sqrt{L/C}$. Identifying that the added capacitance is $C_q$, we obtain
\begin{equation}
    C_q = \Delta C = -\frac{2 \Delta \omega_r}{Z_r \omega_r^2}.
\end{equation}
Using Eq.~(\ref{eq:shift}), yields then the quantum capacitance as
\begin{equation}
    \label{eq:cQEDqCap}
    C_q = \frac{2 (e\alpha)^2 \: t^2 \left< \sigma_z \right>}{\hbar \Gamma (\hbar \omega_q)^2} \frac{x}{1+x^2},
\end{equation}
where $x =2(\omega_q - \omega_r)/\Gamma$. The maximum value of
\begin{equation}
    \label{eq:cQEDqCapMax}
    C_q = \frac{ (e\alpha)^2 \: t^2 \left< \sigma_z \right>}{2\hbar (\omega_q - \omega_r) (\hbar \omega_q)^2},
\end{equation}
is obtained with $x=1$. This result agrees with the semiclassical results of Refs.~\citealp{Petersson2010, Gonzalez-Zalba2016, Chatterjee2018} in the limit $\omega_r \ll \omega_q$. The above JC calculation, however has a peculiar condition for this result: $\Gamma/2 = \omega_q - \omega_r$, which is not present in the semiclassical calculations. The semiclassical derivations consider only the occupancy of the TLS and neglect any coherences that may occur. The results of Eqs.~(\ref{eq:cQEDqCap}) and (\ref{eq:cQEDqCapMax}) take into account the coherent interaction between the readout resonator and the quantum system explicitly. Therefore they contain predictions the decoherence rate $\Gamma$ dependency as well as corrections due to finite resonator frequency $\omega_r$.

\section{Summary}

In summary, \edit{this review considered the Jaynes-Cummings formalism of an electronic two-level system with the photonic mode description. We focused specifically to the transmission response which is typically used in the experiments to probe such systems, and reviewed and connected together several theory concepts that are seen frequently in the experiments.}

\edit{The first part of the review presented the JC system and how it gives rise to the resonance modes with} Lorentzian transmission lineshape. We then discussed the strong coupling regime, $g > \Gamma$, \edit{where the coupling is so strong that the electronic and photonic modes hybridize.} The response \edit{of these hybridized modes} has an analogous form as the \edit{non-hybridized} bare resonator \edit{has. The} input and output coupling and internal losses \edit{of the hybridized modes are obtained as} weighted averages of the \edit{electronic and photonic parts that form the hybridized mode}. The simple \edit{and} intuitive \edit{analytic expressions} provide\edit{d in this review are} a useful tool to determine the underlying coherence of the electronic state based on the \edit{peak transmission} $T(\omega_\pm)$ and the linewidth $\kappa_\pm$ of the hybridized states. The \edit{peak} transmission response considered specifically in this work provides a complementary quantity to use as an input for determining the coherence properties in addition to the usually used linewidth considerations~\cite{Thompson1992, Brune1996, Wallraff2004, Reithmaier2004, Diniz2011, Stockklauser2017, Mi2017, Zeb2018, Delteil2019}. 

\edit{As an important limiting case, we  then considered the case when the TLS energy $\omega_q$ is far detuned from the resonator frequency $\omega_r$. In this case, the mode description results in a frequency shift to the bare resonator response without changing the linewidth. This is the so-called dispersive shift which is used extensively for qubit readout as the frequency shift depends on the qubit state.}

\edit{After the strong coupling regime, the review continued by considering} the weak coupling regime, $g < \Gamma$\edit{, with the same approach.} In this case, the transmission response has only one mode and the TLS gives rise to a frequency shift and \edit{linewidth change} to the mode. We then connected further the frequency shift to the quantum capacitance of a DQD, and the \edit{linewidth change to the underlying} dissipation \edit{that the TLS imposes to the resonator. This dissipation has} the familiar maximum value of $\edit{\kappa}_q = 4g^2/\Gamma$.

\edit{The mode description used in the review maps} the full system to individual modes with their own specific coupling rates\edit{. This provides an additional advantage as it} allows for treating further states hybridizing to these ones iteratively: If another TLS interacts with the system considered here - either via the photonic mode or via the electronic one \edit{-} the Lorentzian mode at either frequency $\omega_\pm$ can be treated as the 'bare' system with which the next TLS hybridizes further. This iterative process is expected to work analogously as long as the other mode $\omega_\mp$ stays sufficiently far away from the response arising from the second hybridization. Such a case can be realized for example at $\delta_q \gtrsim \pm 2g$, where the first hybridization is already taking place partially while keeping the second state sufficiently detuned. 

\section{Acknowledgements}

We thank Pierre Glidic for performing the experiments with double quantum dot states interacting with a high impedance microwave resonator~\cite{Glidic2026} that initiated the consideration on what determines the \edit{peak transmission} and linewidth of an electronic state in the photonic response and the European Union (ERC, QPHOTON, 101087343) and NanoLund for financial support. Views and opinions expressed are however those of the author(s) only and do not necessarily reflect those of the European Union or the European Research Council Executive Agency. Neither the European Union nor the granting authority can be held responsible for them.

\bibliography{main}

@article{Baur2009,
  title = {Measurement of Autler-Townes and Mollow Transitions in a Strongly Driven Superconducting Qubit},
  author = {Baur, M. and Filipp, S. and Bianchetti, R. and Fink, J. M. and G\"oppl, M. and Steffen, L. and Leek, P. J. and Blais, A. and Wallraff, A.},
  journal = {Phys. Rev. Lett.},
  volume = {102},
  issue = {24},
  pages = {243602},
  numpages = {4},
  year = {2009},
  month = {Jun},
  publisher = {American Physical Society},
  doi = {10.1103/PhysRevLett.102.243602},
  url = {https://link.aps.org/doi/10.1103/PhysRevLett.102.243602}
}

@article{Abdumalikov2010,
  title = {Electromagnetically Induced Transparency on a Single Artificial Atom},
  author = {Abdumalikov, A. A. and Astafiev, O. and Zagoskin, A. M. and Pashkin, Yu. A. and Nakamura, Y. and Tsai, J. S.},
  journal = {Phys. Rev. Lett.},
  volume = {104},
  issue = {19},
  pages = {193601},
  numpages = {4},
  year = {2010},
  month = {May},
  publisher = {American Physical Society},
  doi = {10.1103/PhysRevLett.104.193601},
  url = {https://link.aps.org/doi/10.1103/PhysRevLett.104.193601}
}

@article{Astafiev2010,
author = {O. Astafiev  and A. M. Zagoskin  and A. A. Abdumalikov  and Yu. A. Pashkin  and T. Yamamoto  and K. Inomata  and Y. Nakamura  and J. S. Tsai },
title = {Resonance Fluorescence of a Single Artificial Atom},
journal = {Science},
volume = {327},
number = {5967},
pages = {840-843},
year = {2010},
doi = {10.1126/science.1181918},
URL = {https://www.science.org/doi/abs/10.1126/science.1181918},
}

@article{Nilsson2016,
  title = {Single-electron transport in InAs nanowire quantum dots formed by crystal phase engineering},
  author = {Nilsson, Malin and Namazi, Luna and Lehmann, Sebastian and Leijnse, Martin and Dick, Kimberly A. and Thelander, Claes},
  journal = {Phys. Rev. B},
  volume = {93},
  issue = {19},
  pages = {195422},
  numpages = {7},
  year = {2016},
  month = {May},
  publisher = {American Physical Society},
  doi = {10.1103/PhysRevB.93.195422},
  url = {https://link.aps.org/doi/10.1103/PhysRevB.93.195422}
}

@Article{Bjork2002,
author={Bj{\"o}rk, M. T. and Ohlsson, B. J. and Sass, T. and Persson, A. I.
and Thelander, C. and Magnusson, M. H. and Deppert, K. and Wallenberg, L. R.
and Samuelson, L.},
title={One-dimensional Steeplechase for Electrons Realized},
journal={Nano Letters},
year={2002},
month={Feb},
day={01},
publisher={American Chemical Society},
volume={2},
number={2},
pages={87-89},
issn={1530-6984},
doi={10.1021/nl010099n},
url={https://doi.org/10.1021/nl010099n}
}

@article{Clerk2010,
  title = {Introduction to quantum noise, measurement, and amplification},
  author = {Clerk, A. A. and Devoret, M. H. and Girvin, S. M. and Marquardt, Florian and Schoelkopf, R. J.},
  journal = {Rev. Mod. Phys.},
  volume = {82},
  issue = {2},
  pages = {1155--1208},
  numpages = {0},
  year = {2010},
  month = {Apr},
  publisher = {American Physical Society},
  doi = {10.1103/RevModPhys.82.1155},
  url = {https://link.aps.org/doi/10.1103/RevModPhys.82.1155}
}

@ARTICLE{Jaynes1963,
  author={Jaynes, E.T. and Cummings, F.W.},
  journal={Proc. IEEE}, 
  title={Comparison of quantum and semiclassical radiation theories with application to the beam maser}, 
  year={1963},
  volume={51},
  number={1},
  pages={89-109},
  doi={10.1109/PROC.1963.1664}}

@article{Gardiner1984,
  title = {Input and output in damped quantum systems: Quantum stochastic differential equations and the master equation},
  author = {Gardiner, C. W. and Collett, M. J.},
  journal = {Phys. Rev. A},
  volume = {31},
  issue = {6},
  pages = {3761--3774},
  numpages = {0},
  year = {1985},
  month = {Jun},
  publisher = {American Physical Society},
  doi = {10.1103/PhysRevA.31.3761},
  url = {https://link.aps.org/doi/10.1103/PhysRevA.31.3761}
}

@Article{Lindblad1976,
author={Lindblad, G.},
title={On the generators of quantum dynamical semigroups},
journal={Commun. Math. Phys.},
year={1976},
volume={48},
pages={119},
doi={10.1007/BF01608499},
url={https://doi.org/10.1007/BF01608499}
}

@Article{Franke1976,
author={Franke, V. A.},
title={On the general form of the dynamical transformation of density matrices},
journal={Theor. Math. Phys.},
year={1976},
volume={27},
pages={406},
doi={10.1007/BF01051230},
url={https://doi.org/10.1007/BF01051230}
}

@article{Gorini1976,
    author = {Gorini, Vittorio and Kossakowski, Andrzej and Sudarshan, E. C. G.},
    title = {Completely positive dynamical semigroups of N‐level systems},
    journal = {J. Math. Phys.},
    volume = {17},
    number = {5},
    pages = {821-825},
    year = {1976},
    month = {05},
    issn = {0022-2488},
    doi = {10.1063/1.522979},
    url = {https://doi.org/10.1063/1.522979}
}

@Article{Andersson2025,
author={Andersson, S. and Havir, H. and Ranni, A. and Haldar, S. and Maisi, V. F.},
title={High-impedance microwave resonators with two-photon nonlinear effects},
journal={Nature Commun.},
year={2025},
month={Jan},
day={09},
volume={16},
number={1},
pages={552},
issn={2041-1723},
doi={10.1038/s41467-025-55860-8},
url={https://doi.org/10.1038/s41467-025-55860-8}
}

@article{Blais2004,
  title = {Cavity quantum electrodynamics for superconducting electrical circuits: An architecture for quantum computation},
  author = {Blais, Alexandre and Huang, Ren-Shou and Wallraff, Andreas and Girvin, S. M. and Schoelkopf, R. J.},
  journal = {Phys. Rev. A},
  volume = {69},
  issue = {6},
  pages = {062320},
  numpages = {14},
  year = {2004},
  month = {Jun},
  publisher = {American Physical Society},
  doi = {10.1103/PhysRevA.69.062320},
  url = {https://link.aps.org/doi/10.1103/PhysRevA.69.062320}
}

@article{Blais2021,
  title = {Circuit quantum electrodynamics},
  author = {Blais, Alexandre and Grimsmo, Arne L. and Girvin, S. M. and Wallraff, Andreas},
  journal = {Rev. Mod. Phys.},
  volume = {93},
  issue = {2},
  pages = {025005},
  numpages = {72},
  year = {2021},
  month = {May},
  publisher = {American Physical Society},
  doi = {10.1103/RevModPhys.93.025005},
  url = {https://link.aps.org/doi/10.1103/RevModPhys.93.025005}
}

@article{Boissonneault2009,
  title = {Dispersive regime of circuit QED: Photon-dependent qubit dephasing and relaxation rates},
  author = {Boissonneault, Maxime and Gambetta, J. M. and Blais, Alexandre},
  journal = {Phys. Rev. A},
  volume = {79},
  issue = {1},
  pages = {013819},
  numpages = {17},
  year = {2009},
  month = {Jan},
  publisher = {American Physical Society},
  doi = {10.1103/PhysRevA.79.013819},
  url = {https://link.aps.org/doi/10.1103/PhysRevA.79.013819}
}

@article{Brune1996,
  title = {{Quantum Rabi Oscillation: A Direct Test of Field Quantization in a Cavity}},
  author = {Brune, M. and Schmidt-Kaler, F. and Maali, A. and Dreyer, J. and Hagley, E. and Raimond, J. M. and Haroche, S.},
  journal = {Phys. Rev. Lett.},
  volume = {76},
  issue = {11},
  pages = {1800--1803},
  numpages = {0},
  year = {1996},
  month = {Mar},
  publisher = {American Physical Society},
  doi = {10.1103/PhysRevLett.76.1800},
  url = {https://link.aps.org/doi/10.1103/PhysRevLett.76.1800}
}

@Article{Burkard2020,
author={Burkard, Guido and Gullans, Michael J. and Mi, Xiao and Petta, Jason R.},
title={Superconductor--semiconductor hybrid-circuit quantum electrodynamics},
journal={Nature Rev. Phys.},
year={2020},
month={Mar},
day={01},
volume={2},
number={3},
pages={129-140},
issn={2522-5820},
doi={10.1038/s42254-019-0135-2},
url={https://doi.org/10.1038/s42254-019-0135-2}
}

@article{Chatterjee2018,
  title = {A silicon-based single-electron interferometer coupled to a fermionic sea},
  author = {Chatterjee, Anasua and Shevchenko, Sergey N. and Barraud, Sylvain and Otxoa, Rub\'en M. and Nori, Franco and Morton, John J. L. and Gonzalez-Zalba, M. Fernando},
  journal = {Phys. Rev. B},
  volume = {97},
  issue = {4},
  pages = {045405},
  numpages = {7},
  year = {2018},
  month = {Jan},
  publisher = {American Physical Society},
  doi = {10.1103/PhysRevB.97.045405},
  url = {https://link.aps.org/doi/10.1103/PhysRevB.97.045405}
}

@article{Childress2004,
  title = {{Mesoscopic cavity quantum electrodynamics with quantum dots}},
  author = {Childress, L. and S\o{}rensen, A. S. and Lukin, M. D.},
  journal = {Phys. Rev. A},
  volume = {69},
  issue = {4},
  pages = {042302},
  numpages = {8},
  year = {2004},
  month = {Apr},
  publisher = {American Physical Society},
  doi = {10.1103/PhysRevA.69.042302},
  url = {https://link.aps.org/doi/10.1103/PhysRevA.69.042302}
}

@Article{Clerk2020,
author={Clerk, A. A. and Lehnert, K. W. and Bertet, P. and Petta, J. R. and Nakamura, Y.},
title={Hybrid quantum systems with circuit quantum electrodynamics},
journal={Nature Phys.},
year={2020},
month={Mar},
day={01},
volume={16},
number={3},
pages={257-267},
issn={1745-2481},
doi={10.1038/s41567-020-0797-9},
url={https://doi.org/10.1038/s41567-020-0797-9}
}

@Article{Delteil2019,
author={Delteil, Aymeric and Fink, Thomas and Schade, Anne and H{\"o}fling, Sven and Schneider, Christian and {\.{I}}mamo{\u{g}}lu, Ata{\c{c}}},
title={Towards polariton blockade of confined exciton--polaritons},
journal={Nature Materials},
year={2019},
month={Mar},
day={01},
volume={18},
number={3},
pages={219-222},
issn={1476-4660},
doi={10.1038/s41563-019-0282-y},
url={https://doi.org/10.1038/s41563-019-0282-y}
}

@article{Diniz2011,
  title = {Strongly coupling a cavity to inhomogeneous ensembles of emitters: Potential for long-lived solid-state quantum memories},
  author = {Diniz, I. and Portolan, S. and Ferreira, R. and G\'erard, J. M. and Bertet, P. and Auff\`eves, A.},
  journal = {Phys. Rev. A},
  volume = {84},
  issue = {6},
  pages = {063810},
  numpages = {9},
  year = {2011},
  month = {Dec},
  publisher = {American Physical Society},
  doi = {10.1103/PhysRevA.84.063810},
  url = {https://link.aps.org/doi/10.1103/PhysRevA.84.063810}
}

@article{Ferreira2022,
  title = {Microscopic modeling of exciton-polariton diffusion coefficients in atomically thin semiconductors},
  author = {Ferreira, Beatriz and Rosati, Roberto and Malic, Ermin},
  journal = {Phys. Rev. Mater.},
  volume = {6},
  issue = {3},
  pages = {034008},
  numpages = {8},
  year = {2022},
  month = {Mar},
  publisher = {American Physical Society},
  doi = {10.1103/PhysRevMaterials.6.034008},
  url = {https://link.aps.org/doi/10.1103/PhysRevMaterials.6.034008}
}

@article{Frey2012,
  title = {Dipole Coupling of a Double Quantum Dot to a Microwave Resonator},
  author = {Frey, T. and Leek, P. J. and Beck, M. and Blais, A. and Ihn, T. and Ensslin, K. and Wallraff, A.},
  journal = {Phys. Rev. Lett.},
  volume = {108},
  issue = {4},
  pages = {046807},
  numpages = {5},
  year = {2012},
  month = {Jan},
  publisher = {American Physical Society},
  doi = {10.1103/PhysRevLett.108.046807},
  url = {https://link.aps.org/doi/10.1103/PhysRevLett.108.046807}
}

@incollection{Girvin2014,
    author = {Girvin, S. M.},
    isbn = {9780199681181},
    title = "{Circuit QED: superconducting qubits coupled to microwave photons}",
    booktitle = "{Quantum Machines: Measurement and Control of Engineered Quantum Systems: Lecture Notes of the Les Houches Summer School: Volume 96, July 2011}",
    publisher = {Oxford University Press},
    pages = {113-256},
    year = {2014},
    month = {06},
    doi = {10.1093/acprof:oso/9780199681181.003.0003},
    url = {https://doi.org/10.1093/acprof:oso/9780199681181.003.0003}
}

@article{Glidic2026,
    author = {Glidic, P.},
    journal = "{et al. Manuscript in preparation}",
    year = {2026}
}

@Article{Gonzalez-Zalba2016,
author={Gonzalez-Zalba, M. Fernando
and Shevchenko, Sergey N.
and Barraud, Sylvain
and Johansson, J. Robert
and Ferguson, Andrew J.
and Nori, Franco
and Betz, Andreas C.},
title={Gate-Sensing Coherent Charge Oscillations in a Silicon Field-Effect Transistor},
journal={Nano Lett.},
year={2016},
month={Mar},
day={09},
publisher={American Chemical Society},
volume={16},
number={3},
pages={1614-1619},
issn={1530-6984},
doi={10.1021/acs.nanolett.5b04356},
url={https://doi.org/10.1021/acs.nanolett.5b04356}
}

@article{Goppl2008,
    author = {Göppl,M.  and Fragner,A.  and Baur,M.  and Bianchetti,R.  and Filipp,S.  and Fink,J. M.  and Leek,P. J.  and Puebla,G.  and Steffen,L.  and Wallraff,A. },
    title = {Coplanar waveguide resonators for circuit quantum electrodynamics},
    journal = {J. Appl. Phys.},
    volume = {104},
    number = {11},
    pages = {113904},
    year = {2008},
    doi = {10.1063/1.3010859},
    URL = {https://doi.org/10.1063/1.3010859},
}

@book{Haroche2006,
    author = {Haroche, S. and Raimond, J.-M.},
    title = "{{Exploring the Quantum: Atoms, Cavities, and Photons}}",
    publisher = {Oxford University Press},
    year = {2006},
    month = {08},
    isbn = {9780198509141},
    doi = {10.1093/acprof:oso/9780198509141.001.0001},
    url = {https://doi.org/10.1093/acprof:oso/9780198509141.001.0001},
}

@article{Hopfield1958,
  title = {Theory of the Contribution of Excitons to the Complex Dielectric Constant of Crystals},
  author = {Hopfield, J. J.},
  journal = {Phys. Rev.},
  volume = {112},
  issue = {5},
  pages = {1555--1567},
  numpages = {0},
  year = {1958},
  month = {Dec},
  publisher = {American Physical Society},
  doi = {10.1103/PhysRev.112.1555},
  url = {https://link.aps.org/doi/10.1103/PhysRev.112.1555}
}

@article{Houch2008,
  title = {Controlling the Spontaneous Emission of a Superconducting Transmon Qubit},
  author = {Houck, A. A. and Schreier, J. A. and Johnson, B. R. and Chow, J. M. and Koch, Jens and Gambetta, J. M. and Schuster, D. I. and Frunzio, L. and Devoret, M. H. and Girvin, S. M. and Schoelkopf, R. J.},
  journal = {Phys. Rev. Lett.},
  volume = {101},
  issue = {8},
  pages = {080502},
  numpages = {4},
  year = {2008},
  month = {Aug},
  publisher = {American Physical Society},
  doi = {10.1103/PhysRevLett.101.080502},
  url = {https://link.aps.org/doi/10.1103/PhysRevLett.101.080502}
}

@article{Luryi1988,
    author = {Luryi, Serge},
    title = {Quantum capacitance devices},
    journal = {Appl. Phys. Lett.},
    volume = {52},
    number = {6},
    pages = {501-503},
    year = {1988},
    month = {02},
    issn = {0003-6951},
    doi = {10.1063/1.99649},
    url = {https://doi.org/10.1063/1.99649},
}

@Article{Kavokin2015,
author={Kavokin, Alexey V. and Sheremet, Alexandra S. and Shelykh, Ivan A. and Lagoudakis, Pavlos G. and Rubo, Yuri G.},
title={Exciton-photon correlations in bosonic condensates of exciton-polaritons},
journal={Scientific Reports},
year={2015},
month={Jul},
day={08},
volume={5},
number={1},
pages={12020},
issn={2045-2322},
doi={10.1038/srep12020},
url={https://doi.org/10.1038/srep12020}
}

@incollection{Kavokin2017,
    author = {Kavokin, Alexey V. and Baumberg, J. J. and Malpuech, G. and Laussy, F. P.},
    isbn = {9780198782995},
    title = "{Microcavities}",
    publisher = {Oxford University Press},
    year = {2017},
    url = {https://global.oup.com/academic/product/microcavities-9780198782995?cc=fi&lang=en&#}
}

@article{Khan2021,
author={Khan, Waqar
and Potts, Patrick P.
and Lehmann, Sebastian
and Thelander, Claes
and Dick, Kimberly A.
and Samuelsson, Peter
and Maisi, Ville F.},
title={Efficient and continuous microwave photoconversion in hybrid cavity-semiconductor nanowire double quantum dot diodes},
journal={Nature Commun.},
year={2021},
month={Aug},
day={26},
volume={12},
number={1},
pages={5130},
issn={2041-1723},
doi={10.1038/s41467-021-25446-1},
url={https://doi.org/10.1038/s41467-021-25446-1}
}

@article{Kimble1998,
doi = {10.1238/Physica.Topical.076a00127},
url = {https://dx.doi.org/10.1238/Physica.Topical.076a00127},
year = {1998},
month = {jan},
publisher = {},
volume = {1998},
number = {T76},
pages = {127},
author = {H J Kimble},
title = {Strong interactions of single atoms and photons in cavity QED},
journal = {Phys. Scr.},
}

@Article{Kirchmair2013,
author={Kirchmair, Gerhard and Vlastakis, Brian and Leghtas, Zaki and Nigg, Simon E. and Paik, Hanhee and Ginossar, Eran and Mirrahimi, Mazyar and Frunzio, Luigi and Girvin, S. M. and Schoelkopf, R. J.},
title={Observation of quantum state collapse and revival due to the single-photon Kerr effect},
journal={Nature},
year={2013},
month={Mar},
day={01},
volume={495},
number={7440},
pages={205-209},
issn={1476-4687},
doi={10.1038/nature11902},
url={https://doi.org/10.1038/nature11902}
}

@article{Kjaergaard2020,
   author = {Kjaergaard, M. and Schwartz, M. E. and Braumüller, J. and Krantz, P. and Wang, J. I.-J. and Gustavsson, S. and Oliver, W. D.},
   title = {{Superconducting Qubits: Current State of Play}}, 
   journal= {Ann. Rev. Cond. Matt. Phys.},
   year = {2020},
   volume = {11},
   pages = {369-395},
   doi = {https://doi.org/10.1146/annurev-conmatphys-031119-050605},
   url = {https://www.annualreviews.org/content/journals/10.1146/annurev-conmatphys-031119-050605},
}

@Article{FriskKockum2019,
author={Frisk Kockum, A. and Miranowicz, A. and De Liberato, S. and Savasta, S. and Nori, F.},
title={Ultrastrong coupling between light and matter},
journal={Nature Reviews Physics},
year={2019},
month={Jan},
day={01},
volume={1},
number={1},
pages={19-40},
issn={2522-5820},
doi={10.1038/s42254-018-0006-2},
url={https://doi.org/10.1038/s42254-018-0006-2}
}

@article{Mi2017,
author = {X. Mi  and J. V. Cady  and D. M. Zajac  and P. W. Deelman  and J. R. Petta },
title = {{Strong coupling of a single electron in silicon to a microwave photon}},
journal = {Science},
volume = {355},
number = {6321},
pages = {156-158},
year = {2017},
doi = {10.1126/science.aal2469},
URL = {https://www.science.org/doi/abs/10.1126/science.aal2469},
}

@Article{Petersson2010,
author={Petersson, K. D. and Smith, C. G. and Anderson, D. and Atkinson, P. and Jones, G. A. C. and Ritchie, D. A.},
title={Charge and Spin State Readout of a Double Quantum Dot Coupled to a Resonator},
journal={Nano Lett.},
year={2010},
month={Aug},
day={11},
publisher={American Chemical Society},
volume={10},
number={8},
pages={2789-2793},
issn={1530-6984},
doi={10.1021/nl100663w},
url={https://doi.org/10.1021/nl100663w}
}

@article{Ranni2023,
    author = {Ranni, Antti and Havir, Harald and Haldar, Subhomoy and Maisi, Ville F.},
    title = {High impedance Josephson junction resonators in the transmission line geometry},
    journal = {Appl. Phys. Lett.},
    volume = {123},
    number = {11},
    pages = {114002},
    year = {2023},
    month = {09},
    issn = {0003-6951},
    doi = {10.1063/5.0164323},
    url = {https://doi.org/10.1063/5.0164323},
}

@article{Ranni2024,
  title = {Decoherence in a crystal-phase defined double quantum dot charge qubit strongly coupled to a high-impedance resonator},
  author = {Ranni, Antti and Haldar, Subhomoy and Havir, Harald and Lehmann, Sebastian and Scarlino, Pasquale and Baumgartner, Andreas and Sch\"onenberger, Christian and Thelander, Claes and Dick, Kimberly A. and Potts, Patrick P. and Maisi, Ville F.},
  journal = {Phys. Rev. Res.},
  volume = {6},
  issue = {4},
  pages = {043134},
  numpages = {7},
  year = {2024},
  month = {Nov},
  publisher = {American Physical Society},
  doi = {10.1103/PhysRevResearch.6.043134},
  url = {https://link.aps.org/doi/10.1103/PhysRevResearch.6.043134}
}

@article{Reithmaier2004,
author={Reithmaier, J. P. and Sek, G. and L{\"o}ffler, A. and Hofmann, C. and Kuhn, S. and Reitzenstein, S. and Keldysh, L. V. and Kulakovskii, V. D. and Reinecke, T. L. and Forchel, A.},
title={{Strong coupling in a single quantum dot--semiconductor microcavity system}},
journal={Nature},
year={2004},
month={Nov},
day={01},
volume={432},
number={7014},
pages={197-200},
issn={1476-4687},
doi={10.1038/nature02969},
url={https://doi.org/10.1038/nature02969}
}

@article{Scarlino2022,
  title = {In situ Tuning of the Electric-Dipole Strength of a Double-Dot Charge Qubit: Charge-Noise Protection and Ultrastrong Coupling},
  author = {Scarlino, P. and Ungerer, J. H. and van Woerkom, D. J. and Mancini, M. and Stano, P. and M\"uller, C. and Landig, A. J. and Koski, J. V. and Reichl, C. and Wegscheider, W. and Ihn, T. and Ensslin, K. and Wallraff, A.},
  journal = {Phys. Rev. X},
  volume = {12},
  issue = {3},
  pages = {031004},
  numpages = {23},
  year = {2022},
  month = {Jul},
  publisher = {American Physical Society},
  doi = {10.1103/PhysRevX.12.031004},
  url = {https://link.aps.org/doi/10.1103/PhysRevX.12.031004}
}

@article{Stockklauser2017,
  title = {{Strong Coupling Cavity QED with Gate-Defined Double Quantum Dots Enabled by a High Impedance Resonator}},
  author = {Stockklauser, A. and Scarlino, P. and Koski, J. V. and Gasparinetti, S. and Andersen, C. K. and Reichl, C. and Wegscheider, W. and Ihn, T. and Ensslin, K. and Wallraff, A.},
  journal = {Phys. Rev. X},
  volume = {7},
  issue = {1},
  pages = {011030},
  numpages = {6},
  year = {2017},
  month = {Mar},
  publisher = {American Physical Society},
  doi = {10.1103/PhysRevX.7.011030},
  url = {https://link.aps.org/doi/10.1103/PhysRevX.7.011030}
}

@article{Thompson1992,
  title = {{Observation of normal-mode splitting for an atom in an optical cavity}},
  author = {Thompson, R. J. and Rempe, G. and Kimble, H. J.},
  journal = {Phys. Rev. Lett.},
  volume = {68},
  issue = {8},
  pages = {1132--1135},
  numpages = {0},
  year = {1992},
  month = {Feb},
  publisher = {American Physical Society},
  doi = {10.1103/PhysRevLett.68.1132},
  url = {https://link.aps.org/doi/10.1103/PhysRevLett.68.1132}
}

@article{Ungerer2024,
author={Ungerer, J. H. and Pally, A. and Kononov, A. and Lehmann, S. and Ridderbos, J. and Potts, P. P. and Thelander, C. and Dick, K. A. and Maisi, V. F. and Scarlino, P. and Baumgartner, A. and Sch{\"o}nenberger, C.},
title={Strong coupling between a microwave photon and a singlet-triplet qubit},
journal={Nature Commun.},
year={2024},
month={Feb},
day={05},
volume={15},
number={1},
pages={1068},
issn={2041-1723},
doi={10.1038/s41467-024-45235-w},
url={https://doi.org/10.1038/s41467-024-45235-w}
}

@article{vanderWiel2002,
  title = {Electron transport through double quantum dots},
  author = {van der Wiel, W. G. and De Franceschi, S. and Elzerman, J. M. and Fujisawa, T. and Tarucha, S. and Kouwenhoven, L. P.},
  journal = {Rev. Mod. Phys.},
  volume = {75},
  issue = {1},
  pages = {1--22},
  numpages = {0},
  year = {2002},
  month = {Dec},
  publisher = {American Physical Society},
  doi = {10.1103/RevModPhys.75.1},
  url = {https://link.aps.org/doi/10.1103/RevModPhys.75.1}
}

@article{Vigneau2023,
    author = {Vigneau, Florian and Fedele, Federico and Chatterjee, Anasua and Reilly, David and Kuemmeth, Ferdinand and Gonzalez-Zalba, M. Fernando and Laird, Edward and Ares, Natalia},
    title = {Probing quantum devices with radio-frequency reflectometry},
    journal = {Appl. Phys. Rev.},
    volume = {10},
    number = {2},
    pages = {021305},
    year = {2023},
    month = {02},
    issn = {1931-9401},
    doi = {10.1063/5.0088229},
    url = {https://doi.org/10.1063/5.0088229},
}

@article{Wallraff2004,
author={Wallraff, A. and Schuster, D. I. and Blais, A. and Frunzio, L. and Huang, R.- S. and Majer, J. and Kumar, S. and Girvin, S. M. and Schoelkopf, R. J.}, 
title={{Strong coupling of a single photon to a superconducting qubit using circuit quantum electrodynamics}},
journal={Nature},
year={2004},
month={Sep},
day={01},
volume={431},
number={7005},
pages={162-167},
issn={1476-4687},
doi={10.1038/nature02851},
url={https://doi.org/10.1038/nature02851}
}

@article{Yoshie2004,
author={Yoshie, T. and Scherer, A. and Hendrickson, J. and Khitrova, G. and Gibbs, H. M. and Rupper, G. and Ell, C. and Shchekin, O. B. and Deppe, D. G.},
title={{Vacuum Rabi splitting with a single quantum dot in a photonic crystal nanocavity}},
journal={Nature},
year={2004},
month={Nov},
day={01},
volume={432},
number={7014},
pages={200-203},
issn={1476-4687},
doi={10.1038/nature03119},
url={https://doi.org/10.1038/nature03119}
}

@Article{Zeb2018,
author={Zeb, M. Ahsan and Kirton, Peter G. and Keeling, Jonathan},
title={Exact States and Spectra of Vibrationally Dressed Polaritons},
journal={ACS Photonics},
year={2018},
month={Jan},
day={17},
publisher={American Chemical Society},
volume={5},
number={1},
pages={249-257},
doi={10.1021/acsphotonics.7b00916},
url={https://doi.org/10.1021/acsphotonics.7b00916}
}

\end{document}